\documentstyle[12pt]{article}
\newlength{\pubnumber} 

\catcode`\@=11
\@addtoreset{equation}{section}
\def\section{\@startsection{section}{1}{\z@}{3.5ex plus 1ex minus .2ex}
 {2.3ex plus .2ex}{\large\bf}}
\def\subsection{\@startsection{subsection}{2}{\z@}{2.3ex plus .2ex}
 {2.3ex plus .2ex}{\bf}}

\begin{document}

\begin{titlepage}
\samepage{
\setcounter{page}{1}
\rightline{CLNS 92/1168;  McGill/92-41}
\rightline{\tt hep-th/9305093}
\rightline{April 1993}
\vfill
\begin{center}
 {\Large \bf Central Charge Reduction and Spacetime\\
 Statistics in the Fractional Superstring\\}
\vfill
 {\large Philip C. Argyres\footnote{E-mail address:
pca@strange.tn.cornell.edu.  Address after September 1, 1993:  School
of Natural Sciences, Institute for Advanced Study, Princeton, NJ  08540.}\\}
\vspace{.05in}
 {\it  Newman Laboratory of Nuclear Studies\\
Cornell University, Ithaca, NY  14853-5001\\}
\vspace{.15in}
 {\rm and\\}
\vspace{.15in}
 {\large Keith R. Dienes\footnote{E-mail address:
dien@hep.physics.mcgill.ca.}\\}
\vspace{.05in}
 {\it  Department of Physics, McGill University\\
3600 University St., Montr\'eal, Qu\'ebec~H3A-2T8~~Canada\\}
\end{center}
\vfill
\begin{abstract}
 {\rm Fractional superstrings in the tensor-product formulation experience
  ``internal projections'' which reduce their effective central charges.
  Simple expressions for the characters of the resulting effective worldsheet
  theory are found.  All states in the effective theory can be consistently
  assigned definite spacetime statistics.  The projection to the effective
  theory is shown to be described by the action of a dimension-three current
  in the original tensor-product theory.}
\end{abstract}
\vfill}
\end{titlepage}

\setcounter{footnote}{0}

 \font\cmss=cmss10 \font\cmsss=cmss10 at 7pt
\def\IZ{\relax\ifmmode\mathchoice
 {\hbox{\cmss Z\kern-.4em Z}}{\hbox{\cmss Z\kern-.4em Z}}
 {\lower.9pt\hbox{\cmsss Z\kern-.4em Z}}
 {\lower1.2pt\hbox{\cmsss Z\kern-.4em Z}}\else{\cmss Z\kern-.4em Z}\fi}

Fractional superstrings \cite{ALT,AT} have been proposed as a possible
new class of string theories generalizing super- and heterotic strings,
and since then there has been considerable effort in understanding
their worldsheet properties and spacetime phenomenologies
\cite{DT,ADT,ALyT,FL,CR,Scatt}.
These strings have the important property that their critical spacetime
dimensions and central charges are less than those of the superstring,
and this reduction in the critical dimension is accomplished by replacing
the worldsheet supersymmetry of the superstring with a ``$K$-fractional
supersymmetry'' which relates bosons to fields of dimension (spin)
$2/(K+2)$ on the worldsheet, $K\geq 2$.  (The case $K=2$ corresponds to the
ordinary superstring.)
\smallskip

There are two proposals for the identification of the $K$-fractional
worldsheet supersymmetry.  The first proposal \cite{ALT}, which we will
refer to as the chiral algebra approach, associates the fractional
worldsheet supersymmetry with the Virasoro algebra extended via the
inclusion of the fractional supercurrent $G$, a certain chiral operator
of dimension $1+2/(K+2)$.
This algebra is chosen because of its well-behaved representation theory
\cite{ZFft}, and in the $K=4$ case the associated fractional strings
have been shown to have sensible tree-level scattering amplitudes \cite{Scatt}.
Demanding the existence of extra null states in these theories \cite{ALT,ALyT}
indicates that their critical central charges are
\begin{equation}
c_{\rm crit}={6K\over K+2}+{24\over K}.
\label{1}\end{equation}
However, since explicit representations of the $K$-fractional chiral
algebras at these values of the central charge are not known,
one cannot determine the corresponding critical spacetime dimensions
for fractional-superstring propagation.
\smallskip

The second proposal \cite{AT,DT,ADT,ALyT}, which we will refer to as the
tensor-product approach, is based on the observation that the above
$K$-fractional chiral algebras have special representations with central
charges $c_K\equiv 3K/(K+2)$ composed of a free boson plus a
$\IZ_K$-parafermion theory \cite{ZFpf}.  Interpreting the free boson as
a spacetime coordinate of the string, one identifies the $D$-fold tensor
product of the $c_K$ theory with itself as the worldsheet conformal field
theory
(CFT) describing a $K$-fractional string propagating in $D$-dimensional
spacetime.  Since the $K$-fractional chiral algebra is non-linear,
taking tensor products of its representations does not
produce another representaton of the $K$-fractional algebra.  Thus, {\it a
priori}, the two proposals described above for identifying the
$K$-fractional worldsheet supersymmetry are different.  In fact, in the
tensor-product approach, demanding that there be a massless graviton in the
critical dimension and that two dimensions worth of states decouple in
the string propagation (so only ``light-cone'' degrees of freedom
contribute) implies that the critical dimensions are given by
\begin{equation}
D_c=2+{16 \over K}.
\label{2}\end{equation}
This would seem to imply that the critical central charges of these
theories are given by
\begin{equation}
c_{\rm crit}\mathrel{\mathop=^?}D_c\,c_K = {{6K+48}\over{K+2}},
\label{3}\end{equation}
in disagreement with the smaller value in Eq.\ (\ref{1}).
\smallskip

This argument, however, is incorrect.
Anomaly cancellation in the worldsheet theory (in particular, modular
invariance) determines the fractional-superstring partition functions in the
tensor-product approach for the $K=2, 4, 8,$ and 16 cases (corresponding
respectively to the integral critical dimensions $D_c=10,6,4,$ and 3).
Due to delicate cancellations called
``internal projections'' within the characters of the sectors
containing the massless states \cite{ADT,ALyT},
their associated effective central
charges are reduced from the value (\ref{3}).
However, since the partition function only contains information
about physical ({\it i.e.}, light-cone) degrees of freedom,
one can determine in this way only the
central charge of the light-cone CFT.
The result one finds is that the internal projections reduce this
light-cone central charge from $(D-2)c_K=48/(K+2)$ to
\begin{equation}
c_{{\it l.c.}}={24\over K}.
\label{4}\end{equation}
This unexpected reduction in the central charge from its na\"\i ve
tensor-product value indicates that the internal projections reduce the
underlying light-cone CFT from the hypothetical ($D-2$)-fold boson plus
$\IZ_K$-parafermion tensor product to some smaller, non-tensored theory.
Thus, even though the light-cone central charge in Eq.~(\ref{4})
is not directly comparable to the total central charge of Eq.\ (\ref{1}),
it remains a logical possibility
that the chiral algebra and tensor-product approaches to
fractional superstrings are in fact equivalent.
Note, in particular, that restoring $2c_K$ (the na\"\i ve central charge
of the longitudinal and timelike degrees of freedom) to $c_{\it l.c.}$
yields Eq.~(\ref{1}) \cite{ALyT}.
\smallskip

In this letter we focus our attention on the partition functions
derived in the tensor-product approach with the goal of obtaining some
insight into the nature of the internal projections and the properties
of the resulting post-projection light-cone CFT's.  In particular, we
will show how the internal projections act on {\it all}\/ physical
sectors of the tensor-product theory,
and not only those containing the massless
states.  We will also obtain a set of characters
which can be consistently viewed as arising from a (light-cone)
post-projection CFT with the smaller central charge of Eq.\ (\ref{4}).
\smallskip

Let us briefly recall the salient features of the fractional-superstring
partition functions; for more details, see Refs.\ \cite{AT,DT,ADT}.  These
partition functions are composed of combinations of the characters of
$D_c-2$ copies of the free boson plus $\IZ_K$-parafermion theories ---
one for each transverse spacetime dimension.
The characters of each such copy are the so-called ``string
functions''  $c^\ell_n$, labeled
by $SU(2)$ quantum numbers $(\ell/2,n/2)$ with $0\le\ell\le
K$, $\ell\in\IZ$, and $0\le|n|\le\ell$, $n-\ell\in2\IZ$.  (Note that the
string function indices are conventionally twice the usual $SU(2)$
quantum numbers.)  Each $c^\ell_n$ receives contributions from a basic field
$\phi^\ell_n$ or $\phi^\ell_{-n}$ of conformal dimension
\begin{equation}
h^\ell_n={\ell(\ell+2)\over4(K+2)}-{n^2\over4K}\qquad{\rm for}\
|n|\le\ell~,
\label{5}\end{equation}
along with its ``descendent'' fields of dimensions exceeding
$h^\ell_n$ by integers.
One extends the allowed ranges of
$\ell$ and $n$ to arbitrary integers satisfying
$n-\ell\in2\IZ$ via the identifications
$\phi^\ell_n=\phi^\ell_{n+2K}=\phi^{K-\ell}_{n-K}$.
The string functions have power-series expansions in the modular
parameter $q=\exp\{2\pi i\tau\}$ of the form
$c^\ell_n(q)=q^{h^\ell_n-c_K/24}(1+\ldots)$
where the ellipsis indicates terms with positive integral powers of
$q$ and known coefficients \cite{Kac}.
The string functions exhibit the symmetries
$c^\ell_n=c^\ell_{-n}=c^{K-\ell}_{K-n}=c^\ell_{n+2K}$,
and one can show \cite{Gep} that
for any $K$ they transform covariantly under the modular group.
Furthermore, for $K\in 4\IZ$ and $\ell\in 2\IZ$ the combinations
$d^\ell_n\equiv c^\ell_n+c^{K-\ell}_n$
close amongst themselves under modular transformations.
Only these combinations appear in the fractional-superstring
partition functions (see, however, \cite{FL}).
\smallskip

The general form of the fractional-superstring partition function ${\cal
Z}_K$ is then a sum of terms each of which is a product of $D_c-2$
left-moving (holomorphic)
string functions and $D_c-2$ right-moving (anti-holomorphic) string functions.
The coefficients of the terms $q^n{\overline q}^n$ count $(-1)^F$ times
the number of on-shell physical spacetime states of mass-squared $n$,
where $F$ is the spacetime fermion number.
Requiring that the spectrum be tachyon-free,
one finds that the only modular-invariant partition functions
are \cite{AT,ADT}:
\begin{eqnarray}
 {\cal Z}_4&=&({\rm Im}\,\tau)^{-2}\,\left\lbrace|A^b_4-A^f_4|^2+3|B_4|^2
     \right\rbrace, \nonumber\\
 {\cal Z}_8&=&({\rm Im}\,\tau)^{-1}\,\left\lbrace|A^b_8-A^f_8|^2+|B_8|^2+
       2|C_8|^2 \right\rbrace,\nonumber\\
 {\cal Z}_{16}&=&({\rm Im}\,\tau)^{-1/2}\,\left\lbrace|A^b_{16}-A^f_{16}|^2+
    |C_{16}|^2\right\rbrace,
\label{12}\end{eqnarray}
with
\begin{eqnarray}
K=4:~~~~~A^b_4&=&2(d^0_0)^3d^2_0-\textstyle{1\over4}(d^2_0)^4,~~~~~
A^f_4=\textstyle{1\over4}(d^2_2)^4-2(d^0_2)^3d^0_2,\nonumber\\
B_4&=&2(d^0_0)^2d^0_2d^2_2-\textstyle{1\over2}(d^2_0)^2(d^2_2)^2
     +2(d^0_2)^2d^0_0d^2_0,\nonumber\\
K=8:~~~~~A^b_8&=&2d^0_0d^2_0-\textstyle{1\over2}(d^4_0)^2,~~~~~
A^f_8=\textstyle{1\over2}(d^4_4)^2-2d^0_4d^2_4,\nonumber\\
B_8&=&2d^0_0d^2_4-d^4_0d^4_4+2d^2_0d^0_4,\nonumber\\
C_8&=&4d^0_2d^2_2-(d^0_2)^2,\nonumber\\
K=16:~~~~~A^b_{16}&=&d^2_0-\textstyle{1\over2}d^8_0, ~~~~~
 A^f_{16}=\textstyle{1\over2}d^8_8-d^2_8,\nonumber\\
C_{16}&=&2d^2_4-d^8_4.
\label{13}\end{eqnarray}
  It follows from the string-function expansions
that these combinations
each {\it a priori}\/ have $q$-expansions of the forms
$A^{b,f}_K\sim q^0(1+\ldots)$, $B_K\sim q^{1/2}(1+\ldots)$ and $C_K\sim
q^{3/4}(1+\ldots)$, implying that only the $A^{b,f}_K$ sectors
contain massless states.  In a covariant tensor-product approach,
the massless states contributing to the $A^b_K$
characters obey the equations of motion of massless vector bosons in
$D_c$ dimensions, while the massless states of $A^f_K$ are spacetime
spinors \cite{AT,ALyT}.
The relative minus sign between $A^b_K$ and $A^f_K$ in the
partition function, indicative of opposite spacetime statistics, is
consistent with this identification of the massless states.
Furthermore, from the $\IZ_K$ parafermion fusion rules and field
identifications, we see that
the $n$-quantum number is conserved modulo $K$ under fusion, so that
 $A_K^b\times A_K^b=A_K^b$, $A_K^b\times A_K^f=A_K^f$, and
$A_K^f\times A_K^f=A_K^b$.
This supports the identification of the entire sectors
$A^b_K$ and $A^f_K$ as corresponding to spacetime bosons and fermions
respectively.
\smallskip

A novel feature of these $A^{b,f}_K$ combinations, however,
is the appearance of minus signs in their representations in terms
of string functions (\ref{13}).  These signs are not
interpreted as arising from the spacetime statistics factor $(-1)^F$,
and instead indicate a fundamentally new ``internal projection''
which sharply reduces the number of propagating states within each sector.
Rather than project out the entire towers of states associated with the
string functions, however, the internal projections remove only {\it some}\/ of
the states within each tower.  It is straightforward to show that
the $q$-expansions of $A^{b,f}_K$ nevertheless have
only positive integral coefficients, and indeed these remaining
states precisely recombine to fill out the towers
of a new worldsheet CFT.  Specifically, in terms of characters,
one finds \cite{ADT}:
\begin{equation}
A_K^b(\tau)=A_K^f(\tau)=(D_c-2)
\left[{{\vartheta_2(\tau)}\over{2\eta^3(\tau)}}\right]^{(D_c-2)/2},
\label{15}\end{equation}
where $\vartheta_2$ and $\eta$ are the Jacobi and Dedekind functions.
The equality of $A^b_K$ and $A^f_K$ as functions
of $\tau$ demonstrates an exact equality between the numbers of
bosons and fermions at all mass levels in the theory,
suggestive of spacetime supersymmetry.  Furthermore, as functions of
$\tau$, the expressions
$B_K$ and $C_K$ each vanish as well, suggesting that these massive sectors
also enjoy a spacetime supersymmetry \cite{ADT}.
Recognizing $\eta^{-1}$ as the character of a
free worldsheet boson (contributing central charge $c=1$) and
$\sqrt{\vartheta_2/\eta}$ as the character of a worldsheet Majorana fermion
(with central charge $c=1/2$), we see that $A^{b,f}_K$ can be thought of
as characters of an effective (light-cone) CFT with central charge
$c_{{\it l.c.}}={3\over2}(D_c-2)$.
This is the announced reduction of the central charge,
Eq.\ (\ref{4}), from the na\"\i ve tensor-product value.
\smallskip

We would now like to find a splitting of the $B_K$-sectors of the form
$B_K=B^b_K-B^f_K$, realizing the expected spacetime supersymmetry of
 {\it this} sector in terms of explicit boson and fermion characters as well.
Examination of the $B_K$ expressions in Eq.\ (\ref{13}) shows that each
term therein is a product of string functions $d^\ell_n$ in which
half of the $n$-quantum numbers are $0$ and the other half $K/2$.
This means that the simple spin-statistics selection rule
which was used to identify the
$A^b_K$ and $A^f_K$ splitting is not appropriate
for the $B_K$ sectors.  We can, however, attempt to construct combinations
of the string functions occurring in the $B_K$ expressions which
are characters for an effective post-projection light-cone CFT
with the reduced central charge $c_{{\it l.c.}}$ of Eq.\
(\ref{4}).  Indeed, this requirement is necessary for a consistent
light-cone CFT interpretation of both the $A_K$ and $B_K$ sectors of the
fractional-superstring partition functions.  In general, one can deduce
the effective central charge
from a given character $\chi(q)=\sum a_n q^n$ by the asymptotic
behavior \cite{HR} of its coefficients $a_n$ for large $n$:
$a_n\sim \exp(4\pi\sqrt{c_{\rm eff}n/24})$.  Using this criterion,
we find the following unique splitting of the $B_K$:
\begin{eqnarray}
B_4^b &=& 2(d^0_0)^2 d^0_2 d^2_2
- \textstyle{1\over4}(d^2_0)^2(d^2_2)^2,\nonumber\\
B_4^f &=& \textstyle{1\over4}(d^2_2)^2 (d^2_0)^2
- 2(d^0_2)^2 d^0_0 d^2_0,\nonumber\\
B_8^b &=& 2d^0_0 d^2_4 - \textstyle{1\over2}d^4_0 d^4_4,\nonumber\\
B_8^f &=& \textstyle{1\over2}d^4_4 d^4_0 - 2d^0_4 d^2_0.
\label{16}\end{eqnarray}
Moreover, in analogy with the worldsheet boson and fermion characters
found in Eq.~(\ref{15}) for the $A$-sectors, we now find
\begin{equation}
B_K^b(\tau)=B_K^f(\tau)=(D_c -2)
\left[{{\vartheta_2(\lambda\tau)}\over{2\eta^3(\tau)}}\right]^{(D_c-2)/2},
\label{17}\end{equation}
where $\lambda\equiv (h^2_0)^{-1}={1\over2}(K+2)$.
This result can be proven by the
general methods presented in Ref.~\cite{ADT}.
Note that unlike Eq.~(\ref{15}), the argument of the $\vartheta_2$ function
is rescaled by a factor $\lambda$, which for each value of $K$ is
the inverse of the spin of the worldsheet
fractional superpartner of the coordinate boson fields.
\smallskip

We are encouraged to think of $B^{b,f}_K$ as describing
spacetime bosons and fermions because they enter into $B_K$ with a
relative minus sign;  their absolute signs are not determined,
however, so we cannot say definitely which of $B^b_K$ or $B^f_K$ describes
which spacetime statistics.  In the na\"\i ve tensor-product
approach, of course,
the permutation symmetry amongst the different component theories makes it
inconsistent to assign {\it either} bosonic {\it or}\/ fermionic
statistics to $B_K$-sector states \cite{DT,ADT}.  However, with the
understanding that the post-projection light-cone CFT is {\it not}\/ a
tensor product of $D_c-2$ theories each with central charge $c_K$,
we are faced with the possibility that the tensor-product
fusion rules may be further restricted.
One simple example of how this may occur, though by no means
a definitive proposal, is that the internal projection violates the
permutation symmetry between the original tensored theories by rendering the
statistics assignment for the $B_K$-sector states dependent upon the
 {\it ordering}\/ of the $n$-quantum numbers of states in
the tensor product:
\begin{eqnarray}
 {\rm bosonic}\ \Longleftrightarrow &&n_i=
\left( 0,K/2,0,K/2,...,0,K/2\right),\nonumber\\
 {\rm fermionic}\ \Longleftrightarrow &&n_i=
\left( K/2,0,K/2,0,...,K/2,0\right),
\label{18}\end{eqnarray}
where each entry (0 or $K/2$) is repeated $(D_c-2)/2=8/K$ times.
Indeed, this assignment reproduces the correct
statistics selection rules under fusion
not only amongst the $B$-type sectors, but also
between the $A$- and $B$-type sectors.
\smallskip

Given the results (\ref{15}) and (\ref{17}), one can
begin to address the question of determining the fractional-superstring
light-cone conformal field theory {\it after}\/ the internal projection.
Since the $A_K$- and $B_K$-sector splittings are consistent with the reduced
central charges $c_{{\it l.c.}}$ in Eq.~(\ref{4}), it is natural to think of
$A^{b,f}_K$ and $B^{b,f}_K$ as two of the characters of this post-projection
CFT.  Now, the character associated with a primary field of conformal
dimension $h$ in a general CFT of central charge $c$ has
a $q$-expansion of the form $\chi(q)=q^{h-{c/24}}(1+\ldots)$.
Since our post-projection light-cone CFT has the central charge
$c=c_{{\it l.c.}}$,
we can determine the dimensions $h(\chi)$ of the primary fields
associated with these characters, obtaining the results
$h(A^{b,f}_K)=1/K$ and $h(B^{b,f}_K)=\lambda/K$.
In fact one can go even further:  under the
the $S$-modular transformation, the $A^{b,f}_K$ and
$B^{b,f}_K$ characters close on other combinations of string functions
which also contain internal projections
reducing the values of their effective central charges to $c_{{\it l.c.}}$.
In this way we can determine the complete spectrum of highest weights
in these post-projection CFT's, as well as their fusion rules.
A detailed analysis of these post-projection CFT's for both
the $A$- and $B$-sectors is presented in Ref.~\cite{D}.
\smallskip

There exists a simple rule, already apparent from Eqs.\ (\ref{13}) and
(\ref{16}), governing which terms appear with plus and minus signs in
the internally projected characters. Namely, replacing $n\rightarrow
n+K/2$ (mod $K$) in each string function $d^\ell_n$ maps terms occurring
with positive coefficients in $A_K^{b,f}$ and $B_K^{b,f}$
to those with negative coefficients and {\it vice versa}.
In fact, for each relevant value of $K\geq 2$,
there is an operator $\Psi_K$ in the light-cone
tensor-product CFT whose OPE with other fields in that theory
generates this map:
\begin{equation}
\Psi_K \equiv \phi^0_{\pm K/4}\otimes\cdots\otimes\phi^0_{\pm K/4},
\label{20}\end{equation}
where there is one factor $\phi^0_{\pm K/4}$ for each transverse
spacetime dimension.  The action of $\Psi_K$ on
other parafermion fields follows from the $\IZ_K$-parafermion fusion rules
and field identifications.
Interestingly, the conformal dimension of $\Psi_K$ is
$h(\Psi_K)=(D_c-2)h^0_{K/2}=3$, independent of $K$.  The
role played by $\Psi_K$ in effecting these internal projections
and reorganizing the fractional-superstring Fock space
is reminiscent of that played by the ``screening operators'' in the
coulomb gas construction \cite{FF} of the Virasoro minimal
series CFT's from a free boson theory.
Note that $\Psi_K$ was also independently
considered in Ref.\ \cite{CR}, where it served as the basis of an
alternate derivation of the fractional-superstring partition
functions and yielded the splittings (\ref{16}).
\smallskip

Thus far we have not discussed the $C_K$ sectors of the $K=8$ and 16
partition functions.  Whereas the $A_K$ and $B_K$ sectors contained only
fields with $n=0$ or $K/2$, these $C_K$ sectors are formed
from string functions $d^\ell_n$ with $n=K/4$.
We have found that
there are no combinations of $n=K/4$ string functions which have
an internal projection reducing
their effective central charge to $c_{{\it l.c.}}$.  Indeed,
the action of the operator
$\Psi_K:d^\ell_{K/4}\rightarrow d^\ell_{3K/4}=d^\ell_{K/4}$
maps each term in the $C_K$ sectors back to itself.
This implies that the only combination of string
functions in the $C_K$ sectors which suffer an internal projection are the
$C_K$ expressions themselves,
which are identically zero as functions of $q$.
The natural interpretation of this fact is that the internal projections
remove all $C_K$-sector states.  Thus,
the $C_K$'s contain no physical states, and play no role
in the post-projection light-cone CFT.
\smallskip

In summary, we have shown that the fractional-superstring partition
functions obtained from a tensor-product {\it anzatz}\/ for the underlying
CFT have several properties which are not obvious from their expressions
in terms of string functions.  The effective central charges of the
(light-cone) CFT's described by these partition functions are reduced from
their na\"\i ve tensor-product values by the internal projections.  Both the
$A_K$- and $B_K$- sector states can be divided naturally into sets which can be
identified as either spacetime bosons or fermions.  Furthermore, the
characters of these sets have the very simple forms given in Eqs.\
(\ref{15}) and (\ref{17}).  Finally, the internal projections can be described
by the action of a dimension-3 operator in the light-cone tensor-product CFT
which is the product of one $\IZ_K$-parafermion current
$\phi^0_{K/2}$ for each transverse dimension. These unexpected features
provide valuable guides in the search for an explicit construction of
the post-projection light-cone CFT underlying the fractional superstring
partition functions.

\bigskip
\medskip
\leftline{\large\bf Acknowledgments}
\medskip

We are pleased to thank S.-H.H. Tye for many valuable discussions,
and the Aspen Center for Physics for its hospitality during the initial
stages of this work.
This work was supported in part by the National Science Foundation,
the Natural Sciences and Engineering Research Council of Canada,
and les fonds FCAR du Qu\'ebec.

\vfill\eject

\bigskip
\bibliographystyle{unsrt}

\end{document}